\documentclass[11pt]{article}%
\usepackage{amsmath}
\usepackage{amsfonts}
\usepackage{amssymb}
\usepackage{graphicx}
\usepackage{slashed}%
\providecommand{\U}[1]{\protect\rule{.1in}{.1in}}
\newtheorem{theorem}{Theorem}
\newtheorem{acknowledgement}[theorem]{Acknowledgement}

\begin{document}
\bigskip\begin{titlepage}
		\vspace{.3cm} \vspace{1cm}
		\begin{center}
			\baselineskip=16pt
			\centerline{\bf{\Large{Supergravity with mimetic dark matter}}}
			\vspace{1cm}
			\centerline{\large\bf Ali H. Chamseddine }
			\vspace{.5cm}
			\emph{\centerline{Physics Department, American University of Beirut, Lebanon}}
	        \end{center}
		\vspace{1cm}
		\begin{center}
			{\bf Abstract}
		\end{center}
	We formulate a supersymmetric version of gravity with mimetic dark matter. The coupling
of a constrained chiral multiplet to N=1 supergravity is made locally supersymmetric using the
rules of tensor calculus. The chiral multiplet is constrained with a Lagrange multiplier multiplet
that could be either a chiral multiplet or a linear multiplet. We obtain the fully supersymmetric
Lagrangians in both cases. It is then shown that the system consisting of the supergravity multiplet,
the chiral multiplet and the Lagrange multiplier multiplet can break supersymmetry spontaneously
leading to a model of a graviton, massive gravitino and two scalar fields representing mimetic
dark matter. The combination of the chiral multiplet and Lagrange multiplier multiplet can act as the
hidden sector breaking local N=1 supersymmetry.
\end{titlepage}

\section{\bigskip Introduction}

The role of diffeomorphism invariance in General Relativity is to promote
redundancy in presentation of the dynamical degrees of freedom in exchange of
gaining simplicity and elegancy in the formulation. A metric $g_{\mu\nu}$ with
ten components is used to represent two dynamical degrees of freedom for the
graviton. It is then natural to seek a modified theory of gravity without
increasing the degrees of freedom of the system while preserving
diffeomorphism invariance. Some years ago such a modified theory of gravity
was proposed by Mukhanov and myself \cite{mimetic}, based on the idea of
employing a dynamical metric $g_{\mu\nu}$ depending on an auxiliary metric
$\widetilde{g}_{\mu\nu}$ through the relation%
\begin{equation}
g_{\mu\nu}=\widetilde{g}_{\mu\nu}\widetilde{g}^{\kappa\lambda}\partial
_{\kappa}\phi\partial_{\lambda}\phi, \label{scale}%
\end{equation}
where $\phi$ is an additional scalar field. This definition exhibits scale
invariance under the transformation
\begin{equation}
\widetilde{g}_{\mu\nu}\rightarrow\widetilde{g}_{\mu\nu}e^{\omega\left(
x\right)  },
\end{equation}
suggesting a relation between the field $\phi$ and the scale factor of
$g_{\mu\nu}$ as can be seen from the property%
\begin{equation}
g^{\mu\nu}\partial_{\mu}\phi\partial_{\nu}\phi=1 \label{constraint}%
\end{equation}
Equivalently, one can show that the general solution of the constraint
equation (\ref{constraint}) is given by (\ref{scale}). It is easier to impose
the constraint (\ref{constraint}) through a Lagrange multiplier than to work
with an auxiliary metric $\widetilde{g}_{\mu\nu}$ \cite{Golovnev}. It proved
that the simple observation of formulating a theory of gravity subject to the
constraint (\ref{constraint}) has many important consequences. First, it was
shown that \ the system $\left(  g_{\mu\nu},\phi\right)  $\ with eleven fields
constrained by equation (\ref{constraint})\ is equivalent to GR with an
additional half of a degree of freedom for the longitudinal mode that
contributes the equivalent of pressureless dust to the energy-momentum tensor
\cite{mimetic}. Next, it was seen that by adding a potential function
$V\left(  \phi\right)  $ it is possible to construct various physical
cosmological models \cite{mimcos}. In particular, the field $\phi$ could serve
as the inflaton field. In the synchronous gauge a solution of the constraint
equation (\ref{constraint})\ is
\begin{equation}
\phi=t+c, \label{time}%
\end{equation}
and thus the field $\phi$ represents the time coordinate. In general for an
arbitrary metric $g_{\mu\nu}$ the constraint equation (\ref{constraint}%
)\ defines the field $\phi$ as that of synchronous time. For the solution
(\ref{time})\ \ \ we have $\square\phi=\kappa$ where $\kappa$ is the extrinsic
curvature, it was then shown that by adding to the Einstein-Hilbert action
terms of the form $f\left(  \square\phi\right)  $ for some suitable functions
$f,$ it is possible to resolve cosmological singularities \cite{Singular}. The
function $f\left(  \square\phi\right)  $ contributes to the energy momentum
tensor and acts as a source of "mimetic" dark matter providing a geometrical
explanation of dark matter without the need of introducing new forms of
interaction. Exact solutions of the system of equations with a smooth metric
at $t=0$ were found for Friedmann, Kasner universes as well as for black holes
\cite{BH}, \cite{AF}, \cite{minimal}.

On the other hand, supersymmetry, since the 1970's, has played a prominent
role in the attempts to promote gravity to a finite quantum theory within the
settings of supergravity and superstring theory in various dimensions ranging
from four to eleven. Supersymmetry, however, is not observed in nature and
must be spontaneously broken. In most realistic models this is done in the
context of four dimensional $N=1$ supergravity with the gravitino absorbing a
Goldistino fermion \cite{Cremmer79}. The purpose of this article is to
construct a supersymmetric form of gravity with mimetic dark matter. This will
be done using the methods of $N=1$ supersymmetric tensor calculus
\cite{Stelle}, \cite{Cremmer79}, \cite{Cremmer 82}, \cite{ACN1},
\cite{Cremmer83}, \cite{ACN2}. It turns out that there are few number of
possibilities depending on the type of supermultiplet used for the Lagrange
multiplier or the one including the field $\phi.$ In this paper we will
explore two possibilities. In the first, a chiral multiplet is used as a
Lagrange multiplier multiplet to impose a constraint on the metric coupled to
a chiral multiplet. The constraint on the scalar field $z$ is of the form
$z\square z+\overline{z}\square\overline{z}$ instead of $g^{\mu\nu}%
\partial_{\mu}z\partial_{\nu}\overline{z}.$ This is done in section 2. In
section 3 a linear multiplet is used instead as the Lagrange multiplier
multiplet, and results in a constraint on the kinetic term $g^{\mu\nu}%
\partial_{\mu}z\partial_{\nu}\overline{z}$ in line with the non-supersymmetric
case. Section 4 is the
conclusion and section 5 gives the full supersymmetric Lagrangians in
component form for the two cases considered.

\section{Supergravity with mimetic dark matter}

To find the locally superysmmetric form of supergravity with mimetic matter we
have to first specify the nature of the supermultiplets used. In what follows
we will use the notation and conventions of \cite{ACN1}, \cite{ACN2}.\ The
constraint term (\ref{constraint}) is the kinetic part of a real scalar field
can be embedded as part of a left-handed chiral multiplet $\Sigma$ denoted by
\begin{equation}
\Sigma=\left(  z,\chi,h\right)  ,
\end{equation}
with the multiplet of opposite chirality given by
\begin{equation}
\Sigma^{\dagger}=\left(  \overline{z},\chi^{c},\overline{h}\right)  ,
\end{equation}
where $\overline{z},$ $\overline{h}$ are the complex conjugates of $z,$ $h,$
and $\chi^{c}=C^{-1}\overline{\chi}$ is the spinor conjugate to $\chi.$ The
symmetric product $\frac{1}{2}\Sigma^{\dagger}\Sigma$ is a real vector
multiplet of the form
\begin{equation}
V=\left(  C,\zeta,v,V_{\mu},\lambda,D\right)  ,
\end{equation}
where $C,D$ are real scalar fields and $v$ is a complex scalar field,
$\zeta,\lambda$ are Majorana spinors and $V_{\mu}$ is a real vector. To deal
with the most general coupling we consider a function $\phi\left(
\Sigma,\Sigma^{\dagger}\right)  $ of the symmetric products of $\Sigma
^{\dagger}$ and $\Sigma$ whose lowest component is%
\begin{equation}
\phi\left(  z,\overline{z}\right)  =%
{\displaystyle\sum\limits_{m,n}}
a_{m,n}z^{m}\overline{z}^{n}.
\end{equation}
The components of the real vector multiplet are given by
\begin{align}
C  &  =\frac{1}{2}\phi,\quad\zeta=i\left(  \phi_{,z}\chi-\phi_{,\overline{z}%
}\chi^{c}\right)  ,\\
v  &  =-\phi_{,z}h+\phi_{,zz}\overline{\chi^{c}}\chi,\quad V_{\mu}=\frac{i}%
{2}\left(  \phi_{,z}\widehat{D}_{\mu}z-\phi_{,\overline{z}}\widehat{D}_{\mu
}\overline{z}-2\phi_{,z\overline{z}}\overline{\chi}\gamma_{\mu}\chi\right)
,\\
\lambda &  =-i\phi_{,z\overline{z}}\left(  \left(  \overline{h}\chi-h\chi
^{c}\right)  +\widehat{\slashed{D}}\overline{z}\chi-\widehat{\slashed{D}}%
z\chi^{c}\right)  -i\phi_{,zz\overline{z}}\left(  \overline{\chi^{c}}%
\chi\right)  \chi^{c}+i\phi_{,z\overline{z}\overline{z}}\left(  \overline
{\chi}\chi^{c}\right)  \chi,\\
D  &  =\phi_{,z\overline{z}}\left(  \left\vert h\right\vert ^{2}%
-\widehat{D}_{\mu}\overline{z}\widehat{D}^{\mu}z-\overline{\chi}%
\overleftrightarrow{\widehat{\slashed{D}}}\chi\right)  -\phi_{,zz\overline{z}%
}\left(  \overline{h}\overline{\chi^{c}}\chi+\overline{\chi}\gamma^{\mu}%
\chi\widehat{D}_{\mu}z\right) \\
&  -\phi_{,z\overline{z}\overline{z}}\left(  h\overline{\chi}\chi
^{c}+\overline{\chi^{c}}\gamma^{\mu}\chi^{c}\widehat{D}_{\mu}\overline
{z}\right)  +\phi_{,zz\overline{z}\overline{z}}\left(  \overline{\chi^{c}}%
\chi\right)  \left(  \overline{\chi}\chi^{c}\right)  .
\end{align}
We recognize that the $D$-term of the above vector multiplet contains the
kinetic term of the complex scalar field $z$ and for this to be constrained by
a Lagrange multiplier it must couple to a real vector multiplet. The most
economical choice is to use a left-handed chiral multiplet
\begin{equation}
S=\left(  A+iB,\psi,F+iG\right)
\end{equation}
that can be embedded in a real irreducible vector multiplet of the form
\cite{van}
\begin{equation}
\widehat{D}S=\left(  B,\psi,l,\widehat{D}_{\mu}A,0,0\right)
\end{equation}
where $l=F+iG.$ The rule of multiplying two vector multiplets $V=V_{1}\cdot
V_{2}$\ where $V_{1}=\left(  C_{1},\zeta_{1},v_{1},V_{\mu1},\lambda_{1}%
,D_{1}\right)  $\ and \ $V_{2}=\left(  C_{2},\zeta_{2},v_{2},V_{\mu2}%
,\lambda_{2},D_{2}\right)  $\ is given by \cite{Stelle}, \cite{ACN2}
\begin{align}
C  &  =C_{1}C_{2},\quad\zeta=C_{1}\zeta_{2}+C_{2}\zeta_{1},\\
v  &  =C_{1}v_{2}+C_{2}v_{1}-\frac{1}{2}\overline{\zeta}_{1R}\zeta_{2L}%
-\frac{1}{2}\overline{\zeta}_{2R}\zeta_{1L},\\
V_{\mu}  &  =C_{1}V_{\mu2}+C_{2}V_{\mu1}-\frac{i}{2}\overline{\zeta}_{1}%
\gamma_{\mu}\gamma_{5}\zeta_{2},\\
\lambda &  =C_{1}\lambda_{2}-\frac{1}{2}\widehat{\slashed{D}}C_{1}\zeta
_{2}+\frac{1}{2}\widehat{\overline{v}}_{1}\zeta_{2}+\frac{i}{2}\gamma
_{5}\gamma^{\mu}\zeta_{2}V_{\mu1}+1\leftrightarrow2,\\
D  &  =C_{1}D_{2}-\frac{1}{2}\widehat{D}_{\mu}C_{1}\widehat{D}^{\mu}%
C_{2}-\frac{1}{2}V_{\mu1}V_{2}^{\mu}+\frac{1}{2}\overline{v}_{1}%
v_{2}-\overline{\zeta}_{1}\lambda_{2}-\frac{1}{2}\overline{\zeta}%
_{2}\widehat{\slashed{D}}\zeta_{1}+1\leftrightarrow2,
\end{align}
where the various quantities are defined in Reference \cite{ACN2}. The
coupling of minimal $N=1$ supergravity to a vector multiplet is given by the
Lagrangian \cite{Stelle}, \cite{Cremmer79}
\begin{align}
e^{-1}L_{D}  &  =D-\frac{i}{2}\overline{\psi}_{\mu}\gamma_{5}\gamma^{\mu
}\lambda-\frac{1}{3}\left(  \overline{u}v+u\overline{v}\right) \nonumber\\
&  +\frac{2}{3}V_{\mu}\left(  A^{\mu}+\frac{3i}{8}e^{-1}\epsilon^{\mu
\rho\sigma\tau}\overline{\psi}_{\rho}\gamma_{\tau}\psi_{\sigma}\right)
\nonumber\\
&  +\frac{i}{3}e^{-1}\overline{\zeta}\gamma_{5}\gamma_{\mu}R^{\mu}+\frac{i}%
{8}e^{-1}\epsilon^{\mu\nu\rho\sigma}\overline{\psi}_{\mu}\gamma_{\nu}%
\psi_{\rho}\overline{\zeta}\psi_{\sigma}\nonumber\\
&  -\frac{2}{3}e^{-1}C\,L_{SG}, \label{Dterm}%
\end{align}
where
\begin{equation}
e^{-1}L_{SG}=-\frac{1}{2}R\left(  e,\omega\right)  -\frac{1}{3}\left\vert
u\right\vert ^{2}+\frac{1}{3}A_{\mu}A^{\mu}-\frac{1}{2}e^{-1}\overline{\psi
}_{\mu}R^{\mu}. \label{supergravity}%
\end{equation}
The fields $u=S-iP$ and $A_{\mu}$ are auxiliary fields and
\begin{equation}
R^{\mu}=\epsilon^{\mu\nu\rho\sigma}\gamma_{5}\gamma_{\nu}D_{\rho}\left(
\omega\right)  \psi_{\sigma}.
\end{equation}
Examining the components of the product vector multiplet $\frac{1}{2}\left(
\widehat{D}S\cdot\phi\left(  \Sigma,\Sigma^{\dagger}\right)  \right)  $ we
obtain \cite{ACN2}%
\begin{align}
C  &  =\frac{1}{2}B\phi,\quad\xi=iB\left(  \phi_{,z}\chi-\phi_{,\overline{z}%
}\chi^{c}\right)  +\frac{1}{2}\phi\psi,\\
v  &  =-B\left(  h\phi_{,z}-\phi_{,zz}\overline{\chi^{c}}\chi\right)
+\frac{1}{2}\phi l-\frac{i}{2}\overline{\psi}_{R}\left(  \phi_{,z}\chi
_{L}-\phi_{,\overline{z}}\chi_{L}^{c}\right)  +\frac{i}{2}\left(
\phi_{,\overline{z}}\overline{\chi}_{R}-\phi_{,z}\overline{\chi^{c}}%
_{R}\right)  \psi_{L},\\
V_{\mu}  &  =\frac{i}{2}B\left(  \phi_{,z}\widehat{D}_{\mu}z-\phi
_{,\overline{z}}\widehat{D}_{\mu}\overline{z}-2\phi_{,z\overline{z}}%
\overline{\chi}\gamma_{\mu}\chi\right)  +\frac{1}{2}\phi\widehat{D}_{\mu
}A+\frac{1}{2}\overline{\psi}\gamma_{\mu}\gamma_{5}\left(  \phi_{,z}\chi
-\phi_{,\overline{z}}\chi^{c}\right)  ,\\
\lambda &  =-iB\phi_{,z\overline{z}}\left(  \left(  \overline{h}\chi-h\chi
^{c}\right)  +\widehat{\slashed{D}}\overline{z}\chi-\widehat{\slashed{D}}%
z\chi^{c}\right) \nonumber\\
&  -iB\left(  \phi_{,zz\overline{z}}\left(  \overline{\chi^{c}}\chi\right)
\chi^{c}-\phi_{,z\overline{z}\overline{z}}\left(  \overline{\chi}\chi
^{c}\right)  \chi\right)  -\frac{i}{2}\widehat{\slashed{D}}B\left(  \phi
_{,z}\chi-\phi_{,\overline{z}}\chi^{c}\right) \nonumber\\
&  -\frac{1}{4}\widehat{\slashed{D}}\phi\psi-\frac{1}{2}\gamma_{5}\gamma^{\mu
}\left(  \phi_{,z}\chi-\phi_{,\overline{z}}\chi^{c}\right)  \widehat{D}_{\mu
}A-\frac{1}{2}\widehat{\left(  \phi_{,\overline{z}}\overline{h}-\phi
_{,zz}\overline{\chi^{c}}\chi\right)  }\psi\nonumber\\
&  -\frac{1}{4}\gamma_{5}\gamma^{\mu}\psi\left(  \phi_{,z}\widehat{D}_{\mu
}z-\phi_{,\overline{z}}\widehat{D}_{\mu}\overline{z}-2\phi_{,z\overline{z}%
}\overline{\chi}\gamma_{\mu}\chi\right)  +\frac{i}{2}\widehat{\overline{l}%
}\left(  \phi_{,z}\chi-\phi_{,\overline{z}}\chi^{c}\right)  ,\\
D  &  =B\phi_{,z\overline{z}}\left(  \left\vert h\right\vert ^{2}%
-\widehat{D}_{\mu}\overline{z}\widehat{D}^{\mu}z-\overline{\chi}%
\overleftrightarrow{\widehat{\slashed{D}}}\chi\right)  -B\phi_{,zz\overline
{z}}\left(  \overline{h}\overline{\chi^{c}}\chi+\overline{\chi}\gamma^{\mu
}\chi\widehat{D}_{\mu}z\right) \nonumber\\
&  -B\phi_{,z\overline{z}\overline{z}}\left(  h\overline{\chi}\chi
^{c}+\overline{\chi^{c}}\gamma^{\mu}\chi^{c}\widehat{D}_{\mu}\overline
{z}\right)  +B\phi_{,zz\overline{z}\overline{z}}\overline{\chi^{c}}\chi
\cdot\overline{\chi}\chi^{c}\nonumber\\
&  -\frac{1}{2}\widehat{D}^{\mu}B\widehat{D}_{\mu}\phi-\frac{i}{2}%
\widehat{D}^{\mu}A\left(  \phi_{,z}\widehat{D}_{\mu}z-\phi_{,\overline{z}%
}\widehat{D}_{\mu}\overline{z}-2\phi_{,z\overline{z}}\overline{\chi}%
\gamma_{\mu}\chi\right) \nonumber\\
&  -\frac{1}{2}\left(  \overline{l}\left(  h\phi_{,z}-\phi_{,zz}\overline
{\chi^{c}}\chi\right)  +l\left(  \overline{h}\phi_{,\overline{z}}%
-\phi_{,\overline{z}\overline{z}}\overline{\chi}\chi^{c}\right)  \right)
\nonumber\\
&  +i\overline{\psi}\phi_{,z\overline{z}}\left(  \overline{h}\chi-h\chi
^{c}+\widehat{\slashed{D}}\overline{z}\chi-\widehat{\slashed{D}}z\chi
^{c}\right) \nonumber\\
&  +i\left(  \phi_{,zz\overline{z}}\overline{\chi^{c}}\chi\overline{\psi}%
\chi^{c}-\phi_{,z\overline{z}\overline{z}}\overline{\chi}\chi^{c}%
\overline{\psi}\chi\right)  -i\left(  \phi_{,z\overline{z}\overline{z}%
}\overline{\chi}\chi^{c}\overline{\chi}^{c}\psi-\phi_{,zz\overline{z}%
}\overline{\chi^{c}}\chi\overline{\chi}\psi\right) \nonumber\\
&  +\frac{i}{2}\left(  \left(  \phi_{,\overline{z}}\overline{\chi}-\phi
_{,z}\overline{\chi}^{c}\right)  \widehat{\slashed{D}}\psi-\overline{\psi
}\widehat{\slashed{D}}\left(  \phi_{,z}\chi-\phi_{,\overline{z}}\chi
^{c}\right)  \right)  .
\end{align}
We can also add to the action a potential in the form of a holomorphic
function $g\left(  \Sigma\right)  $ of the chiral multiplet $\Sigma$
\cite{Cremmer79}, \cite{ACN2}%
\begin{equation}
g\left(  \Sigma\right)  =\left(  g\left(  z\right)  ,g_{,z}\left(  z\right)
\chi,g_{,z}\left(  z\right)  h-g_{,,zz}\left(  z\right)  \overline{\chi}%
\chi\right)  .
\end{equation}
Coupling of a chiral multiplet $\Sigma$ can be made locally supersymmetric by
extracting the $F$ term in the form according to the formula%
\begin{equation}
e^{-1}L_{F}=\operatorname{Re}\left(  h+uz+\overline{\psi}_{\mu}\gamma^{\mu
}\chi+\overline{\psi}_{\mu}\sigma^{\mu\nu}\psi_{\nu R}z\right)  .
\end{equation}
Applying this to the potential function $g\left(  \Sigma\right)  $ gives
\cite{Cremmer79}, \cite{ACN2}
\begin{equation}
e^{-1}L_{F}=\frac{1}{2}\left(  g_{,z}h-g_{,,zz}\overline{\chi}\chi
+ug+g_{,z}\overline{\psi}_{\mu}\gamma^{\mu}\chi+g\overline{\psi}_{\mu}%
\sigma^{\mu\nu}\psi_{\nu R}+h.c\right)  . \label{Fterm}%
\end{equation}
The total action is then that of supergravity coupled to a constrained chiral
multiplet $\Sigma$ with potential $g\left(  \Sigma\right)  $ through a
Lagrange multiplier chiral multiplet $S$ with an action%
\begin{equation}
e^{-1}L=e^{-1}L_{SG}+e^{-1}L_{D}\left(  S.\phi\left(  \Sigma,\overline{\Sigma
}\right)  \right)  +e^{-1}L_{F}\left(  g\left(  \Sigma\right)  \right)  .
\label{fullS}%
\end{equation}
The full Lagrangian in component form is given in the appendix equation
(\ref{LagrangianA}).

At this point, in order to compare with mimetic gravity, we will analyze the
bosonic part of the action. After integrating by parts the $\partial_{\mu}B$
term, we get%
\begin{align*}
e^{-1}L^{\mathrm{bosonic}}  &  =B\phi_{,z\overline{z}}\left\vert h\right\vert
^{2}+\frac{1}{2}Bg^{\mu\nu}\left(  \phi_{,zz}\partial_{\mu}z\partial_{\nu
}z+\phi_{,\overline{z}\overline{z}}\partial_{\mu}\overline{z}\partial_{\nu
}\overline{z}\right)  -\frac{i}{2}\partial^{\mu}A\left(  \phi_{,z}%
\partial_{\mu}z-\phi_{,\overline{z}}\partial_{\mu}\overline{z}\right) \\
&  -\frac{1}{2}\left(  \overline{l}h\phi_{,z}+l\overline{h}\phi_{,\overline
{z}}\right)  +\frac{1}{3}A^{\mu}\left(  iB\left(  \phi_{,z}\partial_{\mu
}z-\phi_{,\overline{z}}\partial_{\mu}\overline{z}\right)  +\phi\partial_{\mu
}A\right) \\
&  +\left(  \frac{1}{3}B\phi-1\right)  \left(  \frac{1}{2}R+\frac{1}%
{3}\left\vert u\right\vert ^{2}-\frac{1}{3}A_{\mu}A^{\mu}\right) \\
&  -\frac{1}{3}\left(  \overline{u}\left(  -Bh\phi_{,z}+\frac{1}{2}\phi
l\right)  +u\left(  -B\overline{h}\phi_{,\overline{z}}+\frac{1}{2}%
\phi\overline{l}\right)  \right) \\
&  +\frac{1}{2}\left(  g_{,z}h+ug+\overline{g}_{,\overline{z}}\overline
{h}+\overline{u}\overline{g}\right)  .
\end{align*}
Integrating by parts and solving the auxiliary field equations for $A,B,l,h,$
and $u$ we get a coupled system of equations. We start first with the
equations of the Lagrange multiplier multiplet $S.$ The $B$ equation gives
\begin{align}
0  &  =\phi_{,z\overline{z}}\left\vert h\right\vert ^{2}+\frac{1}{2}g^{\mu\nu
}\left(  \phi_{,zz}\partial_{\mu}z\partial_{\nu}z+\phi_{,\overline{z}%
\overline{z}}\partial_{\mu}\overline{z}\partial_{\nu}\overline{z}\right)
+\frac{i}{3}A^{\mu}\left(  \phi_{,z}\partial_{\mu}z-\phi_{,\overline{z}%
}\partial_{\mu}\overline{z}\right) \nonumber\\
&  +\frac{1}{3}\phi\left(  \frac{1}{2}R+\frac{1}{3}\left\vert u\right\vert
^{2}-\frac{1}{3}A_{\mu}A^{\mu}\right)  +\frac{1}{3}\left(  \overline{u}%
h\phi_{,z}+u\overline{h}\phi_{,\overline{z}}\right)  . \label{Bequation}%
\end{align}
The $A$ equation is%
\begin{equation}
0=\frac{i}{2}\partial^{\mu}\left(  \phi_{,z}\partial_{\mu}z-\phi
_{,\overline{z}}\partial_{\mu}\overline{z}\right)  -\frac{1}{3}\partial^{\mu
}\left(  \phi A_{\mu}\right)  , \label{Aequation}%
\end{equation}
while the $\overline{l}$ equation gives
\begin{equation}
0=h\phi_{,z}+\frac{1}{3}u\phi. \label{lequation}%
\end{equation}
The equation for the auxiliary field $\overline{h}$ of the mimetic multiplet
$\Sigma^{\dagger}$ gives%
\begin{equation}
0=Bh\phi_{,z\overline{z}}-\frac{1}{2}l\phi_{,\overline{z}}+\frac{1}{3}%
uB\phi_{,\overline{z}}+\frac{1}{2}\overline{g}_{\overline{z}}.
\label{hequation}%
\end{equation}
Finally, the equations for the supergravity auxiliary fields $\overline{u}$
and $A^{\mu}$ are
\begin{align}
0  &  =\frac{1}{3}\left(  \frac{B\phi}{3}-1\right)  u-\frac{1}{6}\phi
l+\frac{1}{3}Bh\phi_{,z}+\frac{1}{2}\overline{g},\label{uequation}\\
0  &  =\left(  iB\left(  \phi_{,z}\partial_{\mu}z-\phi_{,\overline{z}}%
\partial_{\mu}\overline{z}\right)  +\phi\partial_{\mu}A\right)  -2\left(
\frac{B\phi}{3}-1\right)  A_{\mu}. \label{Amuequation}%
\end{align}
We now solve this system of equations in terms of the complex mimetic field
$z.$ The auxiliary fields $u,$ $A_{\mu},$ $h,$ $l$ are then given by%
\begin{align}
l  &  =3\frac{\overline{g}}{\phi}+6h\frac{\phi_{,z}}{\phi^{2}},\quad
u=-3h\frac{\phi_{,z}}{\phi},\\
h  &  =\frac{\frac{3\overline{g}}{\phi}-\frac{\overline{g}_{,\overline{z}}%
}{\phi_{,\overline{z}}}}{2\left(  B\left(  \frac{\phi_{,z\overline{z}}}%
{\phi_{,\overline{z}}}-\frac{\phi_{,z}}{\phi}\right)  -3\frac{\phi_{,z}}%
{\phi^{2}}\right)  },\\
A_{\mu}  &  =\frac{1}{2\left(  \frac{B\phi}{3}-1\right)  }\left(  iB\left(
\phi_{,z}\partial_{\mu}z-\phi_{,\overline{z}}\partial_{\mu}\overline
{z}\right)  +\phi\partial_{\mu}A\right)  .
\end{align}
Substituting these values into equations (\ref{Bequation}) and
(\ref{Aequation}) as well as using the $z$ equations of motion, we get
differential equations sufficient to determine the fields $A,B$ and $z$ .
However, unlike gravity with mimetic field, the auxiliary fields in
supergravity, $u$ and $A_{\mu}$ do contribute to the effective action and
would cause both the phase factor of the field $z$ as well as the field $A$ to
propagate implying that the bosonic sector will have two degrees of freedom
for the graviton and two degrees of freedom for the scalar fields. The
solutions depend on the choice of functions $\phi\left(  z,\overline
{z}\right)  $ and $g\left(  z\right)  $ and for reasonable choices of these
functions the equations will simplify. In particular we can use a linear
potential
\begin{equation}
g\left(  z\right)  =m^{2}\left(  z+\beta\right)  ,
\end{equation}
the same form used for the hidden sector breaking local supersymmetry
spontaneously \cite{Cremmer79}, \cite{ACN1}. Similarly we can use a simple
choice for the function $\phi\left(  z,\overline{z}\right)  $ such as
\begin{equation}
\phi\left(  z,\overline{z}\right)  =\phi_{0}\exp\left(  -c\overline
{z}z\right)  ,
\end{equation}
which would greatly simplify the equations. We notice that the $B$ equation
(\ref{Bequation}) gives a constraint on the combination $g^{\mu\nu}\left(
\phi_{,zz}\partial_{\mu}z\partial_{\nu}z+\phi_{,\overline{z}\overline{z}%
}\partial_{\mu}\overline{z}\partial_{\nu}\overline{z}\right)  $, instead of
$g^{\mu\nu}\phi_{,z\overline{z}}\partial_{\mu}z\partial_{\nu}\overline{z}.$
This should not be necessarily a problem because the metric $g^{\mu\nu}$ has a
Lorentzian signature and derivatives with respect to time and space
coordinates come with opposite signs. Thus constraints of this type could
always be satisfied by allowing the real and imaginary parts to depend on both
time and space coordinates. This is a situation encountered in our
construction of inhomogeneous dark energy \cite{inhomogeneous}. In the
fermionic sector the Lagrange multiplier fermion $\psi$ constrains the mimetic
field $\chi$ which does couple to the gravitino $\psi_{\mu}.$ A careful
Hamiltonian analysis along the lines of \cite{Ola} is needed to show
explicitly which combination of fields propagate and which ones correspond to
mimetic dark matter in the form of two scalars. The structure of the action
indicates that one can use the potential function $g\left(  z\right)  $ to
break supersymmetry. In this case the singlet field $\Sigma$ does act as the
hidden sector in supergravity triggering the breakdown of supersymmetry
\cite{Cremmer79}, \cite{ACN1}. In the absence of supersymmetry breaking the
fermionic degrees of freedom match the bosonic ones and one can see from the
action of the constrained fermionic system $\left(  \psi,\chi\right)  $ that a
massless chiral fermion propagate. When supersymmetry is broken spontaneously
the massless fermion will be absorbed by the gravitino which becomes massive
with four degrees of freedom matching the four degrees of the graviton plus
two scalars. We will further elaborate on this in section 4 when we consider
spontaneous breaking of supersymmetry.

\section{Real linear multiplet as Lagrange multiplier}

A real linear multiplet is real vector superfield satisfying the constraint
\begin{equation}
D^{2}L=0=\overline{D}^{2}L.
\end{equation}
This implies that the surviving components of the vector multiplet take the
form \cite{derendinger}%
\begin{equation}
L=\left(  A,\psi,b_{\mu},0,-\widehat{\slashed{D}}\psi,-\square^{P}A\right)  ,
\end{equation}
where $b_{\mu}$ is a divergence free vector $\partial^{\mu}b_{\mu}=0.$ We now
couple the linear multiplet $L$ to the symmetric product function $\phi\left(
\Sigma,\Sigma^{\dagger}\right)  $ to obtain a vector multiplet $V$ with
components%
\begin{align}
C  &  =\frac{1}{2}A\phi,\quad\zeta=iA\left(  \phi_{,z}\chi-\phi_{,\overline
{z}}\chi^{c}\right)  +\frac{1}{2}\phi\psi,\\
v  &  =-A\left(  \phi_{,z}h-\phi_{,zz}\overline{\chi^{c}}\chi\right)
-\frac{i}{2}\overline{\psi}_{R}\left(  \phi_{,z}\chi_{L}-\phi_{,\overline{z}%
}\chi_{L}^{c}\right)  +\frac{i}{2}\left(  \phi_{,\overline{z}}\overline{\chi
}_{R}-\phi_{,z}\overline{\chi^{c}}_{R}\right)  \psi_{L},\\
V_{\mu}  &  =\frac{i}{2}A\left(  \phi_{,z}\widehat{D}_{\mu}z-\phi
_{,\overline{z}}\widehat{D}_{\mu}\overline{z}-2\phi_{,z\overline{z}}%
\overline{\chi}\gamma_{\mu}\chi\right)  +\frac{1}{2}\phi b_{\mu}+\frac{1}%
{2}\overline{\psi}\gamma_{\mu}\gamma_{5}\left(  \phi_{,z}\chi-\phi
_{,\overline{z}}\chi^{c}\right)  ,\\
\lambda &  =A\phi_{,z\overline{z}}\left(  -i\overline{h}\chi+ih\chi
^{c}-i\widehat{\slashed{D}}\overline{z}\chi+i\widehat{\slashed{D}}z\chi
^{c}\right)  -\frac{1}{2}\phi\widehat{\slashed{D}}\psi\nonumber\\
&  -\frac{i}{2}\widehat{\slashed{D}}A\left(  \phi_{,z}\chi-\phi_{,\overline
{z}}\chi^{c}\right)  -\frac{1}{4}\widehat{\slashed{D}}\phi\psi-iA\left(
\phi_{,zz\overline{z}}\overline{\chi^{c}}\chi\chi^{c}-\phi_{,z\overline
{z}\overline{z}}\overline{\chi}\chi^{c}\chi\right) \nonumber\\
&  -\frac{1}{2}\gamma_{5}\gamma^{\mu}\left(  b_{\mu}\left(  \phi_{,z}\chi
-\phi_{,\overline{z}}\chi^{c}\right)  +\frac{1}{2}\left(  \phi_{,z}%
\widehat{D}_{\mu}z-\phi_{,\overline{z}}\widehat{D}_{\mu}\overline{z}%
-2\phi_{,z\overline{z}}\overline{\chi}\gamma_{\mu}\chi\right)  \psi\right)
\nonumber\\
&  -\widehat{\frac{1}{2}\left(  \phi_{,\overline{z}}\overline{h}%
-\phi_{,\overline{z}\overline{z}}\overline{\chi}\chi^{c}\right)  }\psi,\\
D  &  =A\phi_{,z\overline{z}}\left(  \left\vert h\right\vert ^{2}%
-\widehat{D}_{\mu}\overline{z}\widehat{D}^{\mu}z-\overline{\chi}%
\overleftrightarrow{\widehat{\slashed{D}}}\chi\right)  -A\phi_{,zz\overline
{z}}\left(  h^{\ast}\overline{\chi^{c}}\chi+\overline{\chi}\gamma^{\mu}%
\chi\widehat{D}_{\mu}z\right) \nonumber\\
&  -A\phi_{,z\overline{z}\overline{z}}\left(  h\overline{\chi}\chi
^{c}+\overline{\chi^{c}}\gamma^{\mu}\chi^{c}\widehat{D}_{\mu}\overline
{z}\right)  +A\phi_{,zz\overline{z}\overline{z}}\overline{\chi^{c}}\chi
\cdot\overline{\chi}\chi^{c}-\frac{1}{2}\phi\square^{P}A-\frac{1}%
{2}\widehat{D}^{\mu}A\widehat{D}_{\mu}\phi\nonumber\\
&  -\frac{i}{2}b^{\mu}\left(  \phi_{,z}\widehat{D}_{\mu}z-\phi_{,\overline{z}%
}\widehat{D}_{\mu}\overline{z}-2\phi_{,z\overline{z}}\overline{\chi}\gamma
_{5}\gamma_{\mu}\chi\right)  +i\phi_{,z\overline{z}}\overline{\psi}\left(
\overline{h}\chi-h\chi^{c}+\widehat{\slashed{D}}\overline{z}\chi
-\widehat{\slashed{D}}z\chi^{c}\right) \nonumber\\
&  -i\phi_{,z\overline{z}}\left(  h\overline{\chi}-\overline{h}\overline
{\chi^{c}}+\widehat{\slashed{D}}z\overline{\chi}-\widehat{\slashed{D}}%
\overline{z}\overline{\chi^{c}}\right)  \psi+\frac{i}{2}\left(  \phi
_{,\overline{z}}\overline{\chi}-\phi_{,z}\overline{\chi^{c}}\right)
\widehat{\slashed{D}}\psi\nonumber\\
&  -\frac{i}{2}\overline{\psi}\widehat{\slashed{D}}\left(  \phi_{,z}\chi
-\phi_{,\overline{z}}\chi^{c}\right)  +i\left(  \phi_{,zz\overline{z}}\left(
\overline{\chi^{c}}\chi\right)  \overline{\psi}\chi^{c}-\phi_{,z\overline
{z}\overline{z}}\left(  \overline{\chi}\chi^{c}\right)  \overline{\psi}%
\chi\right) \nonumber\\
&  -i\left(  \phi_{,z\overline{z}\overline{z}}\overline{\chi}\chi^{c}%
\overline{\chi}^{c}\psi-\phi_{,zz\overline{z}}\overline{\chi^{c}}\chi
\overline{\chi}\psi\right)  .
\end{align}
This multiplet is coupled to supergravity using the formula (\ref{Dterm}). In
addition we add the potential term $g\left(  \Sigma\right)  $ using formula
(\ref{Fterm}). The full Lagrangian in this case is given by%
\begin{equation}
e^{-1}L=e^{-1}L_{SG}+e^{-1}L_{D}\left(  L.\phi\left(  \Sigma,\overline{\Sigma
}\right)  \right)  +e^{-1}L_{F}\left(  g\left(  \Sigma\right)  \right)  .
\label{FullB}%
\end{equation}
The component form of this action is given in the appendix equation
(\ref{LagrangianB}). Collecting all bosonic terms we obtain the Lagrangian%
\begin{align}
e^{-1}L^{\mathrm{bosonic}}  &  =A\phi_{,z\overline{z}}\left(  \left\vert
h\right\vert ^{2}-\partial_{\mu}\overline{z}\partial^{\mu}z\right)  -\frac
{i}{2}b^{\mu}\left(  \phi_{,z}\partial_{\mu}z-\phi_{,\overline{z}}%
\partial_{\mu}\overline{z}\right) \nonumber\\
&  +\frac{A}{3}\left(  \overline{u}\phi_{,z}h+u\phi_{,\overline{z}}%
\overline{h}\right)  +\frac{1}{3}A^{\mu}\left(  iA\left(  \phi_{,z}%
\partial_{\mu}z-\phi_{,\overline{z}}\partial_{\mu}\overline{z}\right)  +\phi
b_{\mu}\right) \nonumber\\
&  +\left(  \frac{1}{3}A\phi-1\right)  \left(  \frac{1}{2}R+\frac{1}%
{3}\left\vert u\right\vert ^{2}-\frac{1}{3}A_{\mu}A^{\mu}\right)  +\frac{1}%
{2}\left(  g_{,z}h+ug+\overline{g}_{,\overline{z}}\overline{h}+\overline
{u}\overline{g}\right)
\end{align}
where we have integrated by parts the $\widehat{D}^{\mu}A$ term to cancel the
$\square^{P}A$ term leaving only the $A\widehat{D}_{\mu}\overline
{z}\widehat{D}^{\mu}z$ kinetic term just as in gravity with mimetic matter. We
first vary the Lagrange multiplier $A$ to obtain%
\begin{align}
0  &  =\phi_{,z\overline{z}}\left(  \left\vert h\right\vert ^{2}-\partial
_{\mu}\overline{z}\partial^{\mu}z\right)  +\frac{1}{3}\left(  \overline{u}%
\phi_{,z}h+u\phi_{,\overline{z}}\overline{h}\right)  +\frac{i}{3}A^{\mu
}\left(  \phi_{,z}\partial_{\mu}z-\phi_{,\overline{z}}\partial_{\mu}%
\overline{z}\right) \nonumber\\
&  +\frac{1}{3}\phi\left(  \frac{1}{2}R+\frac{1}{3}\left\vert u\right\vert
^{2}-\frac{1}{3}A_{\mu}A^{\mu}\right)  \label{linearA}%
\end{align}
Next we vary the divergence free vector $b_{\mu}$ to obtain%
\begin{equation}
0=\frac{i}{2}\left(  \phi_{,z}\partial_{\mu}z-\phi_{,\overline{z}}%
\partial_{\mu}\overline{z}\right)  -\frac{1}{3}\left(  \phi A_{\mu}\right)
+\partial_{\mu}\alpha
\end{equation}
where $\alpha$ is determined from transversality of $b_{\mu}.$ We next vary
the auxiliary field $\overline{h}$ of the chiral multiplet $\Sigma^{\dagger}$
to get%
\begin{equation}
0=A\left(  \phi_{,z\overline{z}}h+\frac{1}{3}u\phi_{,\overline{z}}\right)
+\frac{1}{2}\overline{g}_{,\overline{z}}%
\end{equation}
Finally we vary the supergravity auxiliary fields $\overline{u}$ and $A^{\mu}$
to obtain, respectively%
\begin{align}
0  &  =\frac{1}{3}\left(  \frac{1}{3}A\phi-1\right)  u+\frac{A}{3}\phi
_{,z}h+\frac{1}{2}\overline{g}\\
0  &  =\left(  iA\left(  \phi_{,z}\partial_{\mu}z-\phi_{,\overline{z}}%
\partial_{\mu}\overline{z}\right)  +\phi b_{\mu}\right)  -2\left(  \frac{1}%
{3}A\phi-1\right)  A_{\mu}%
\end{align}
These equations allow us to solve for $u,$ $A_{\mu}$, $h$ and $b_{\mu}$ in
terms of $A$ and $z$ which are constrained by equations (\ref{linearA}) and
the $z$ equation. These are given by%
\begin{align}
h  &  =-\frac{1}{2\left(  \frac{A}{3}\left(  \phi\phi_{,z\overline{z}}%
-\phi_{,z}\phi_{,\overline{z}}\right)  -\phi_{,z\overline{z}}\right)  }\left(
\left(  \frac{\phi}{3}-\frac{1}{A}\right)  g_{,\overline{z}}^{\ast}%
-\phi_{,\overline{z}}g^{\ast}\right) \\
u  &  =\frac{1}{2\left(  \frac{A}{3}\left(  \phi\phi_{,z\overline{z}}%
-\phi_{,z}\phi_{,\overline{z}}\right)  -\phi_{,z\overline{z}}\right)  }\left(
\phi_{,z}g_{,\overline{z}}^{\ast}-3\phi_{,z\overline{z}}g^{\ast}\right)
\label{linearu}\\
A_{\mu}  &  =\frac{3}{\phi}\partial_{\mu}\alpha+\frac{3i}{2\phi}\left(
\phi_{,z}\partial_{\mu}z-\phi_{,\overline{z}}\partial_{\mu}\overline{z}\right)
\label{linearamu}\\
b_{\mu}  &  =\left(  \frac{A}{\phi}-\frac{3}{\phi^{2}}\right)  \partial_{\mu
}\alpha-\frac{3i}{\phi}\left(  \phi_{,z}\partial_{\mu}z-\phi_{,\overline{z}%
}\partial_{\mu}\overline{z}\right)
\end{align}
Noting that the Lagrange multiplier multiplet couples linearly, all terms in
such couplings drop out from the action leaving only the supergravity action
\begin{equation}
-\left(  \frac{e}{2}R+\frac{e}{3}\left\vert u\right\vert ^{2}-\frac{e}%
{3}A_{\mu}A^{\mu}+\frac{1}{2}\overline{\psi}_{\mu}R^{\mu}\right)
\end{equation}
where $u$ and $A_{\mu}$ are given in equations (\ref{linearu}) and
(\ref{linearamu}). The field $A$ will be determined from the components of the
Einstein tensor $G_{\mu\nu}.$ Again we obtain four dynamical degrees of
freedom, two for the graviton and two for the scalar fields $\alpha$ and
imaginary part of $z.$

\section{Conclusions and road map to future work}

Gravity with mimetic dark matter was shown to be a versatile model solving
many outstanding problems. To name few of the advantages we mention first the
generation of mimetic dark matter in the form of dust or geometric extrinstic
curvature contributions to the energy-momentum tensor \cite{mimetic},
\cite{Singular}. Including appropriate functions $f\left(  \square\phi\right)
$ in the action lead to non-singular cosmological and black hole solutions.
Also, with the aid of the mimetic field $\phi$ it is possible to construct
Horava type models of renormalizable gravity without breaking diffeomorphism
invariance \cite{Horava}. All these gains were achieved without introducing
new degrees of freedom to the graviton, except for half a degree corresponding
to the longitudinal mode. On the other hand supersymmetry have attracted a lot
of attention and is looked at as the most viable theory beyond the Standard
Model. In particular $N=1$ locally supersymmetric theories can result as the
low energy limit of consistent superstring models. As supersymmetry is not
observed in nature it must be broken spontaneously. This is usually done
through a hidden sector represented by a singlet chiral supermultiplet
\cite{Cremmer79}, \cite{ACN1}. It is then natural to combine both ideas and
ask the question whether it is possible to supersymmetrize the action of
gravity with mimetic dark matter. Although the gravity action was first
constructed by maintaining the symmetry (\ref{scale}), it proved easier to
equivalently impose the constraint equation(\ref{constraint}) \cite{Golovnev}.
In this work we imposed a constraint on the symmetric product $\Sigma
^{\dagger}\Sigma$ of a chiral supermultiplet $\Sigma=\left(  z,\chi,h\right)
$ with its conjugate $\Sigma^{\dagger}$ first using a chiral supermultiplet
$S=\left(  A+iB,\psi,l\right)  $ as a Lagrange multiplet and later using
instead a linear multiplet $L=\left(  C,\psi,b_{\mu}\right)  .$ We have
constructed the most general coupling of this system to supergravity
represented by the multiplet $SG=\left(  e_{\mu}^{a},\psi_{\mu},u,A_{\mu
}\right)  .$ 

We have shown that the mimetic dark matter supermultiplet do couple to the
auxiliary fields of supergravity. These will then contribute to the action of
auxiliary fields and in general will have non-trivial solutions. In addition,
the gravitino couples to the superpotential and for particular solutions for
the field $z$ satisfying the constraint equations, preliminary work strongly
indicates that supersymmetry will be broken spontaneously. The aim of future
work is to study that the system $S,$ $\Sigma,$ $SG$ or $L,$ $\Sigma,$ $SG$ in
presence of a superpotential $g\left(  z\right)  $, and to determine the
spectrum. In particular a careful analysis is needed to check how the
gravitino field $\psi_{\mu}$ absorbs the Goldstino field. It is known that in
generic models of local supersymmetry  a singlet field with appropriate
superpotential function $g\left(  z\right)  $ can break supersymmetry
spontaneously and that the gravitino field absorbs the Goldistino to become
massive \cite{Cremmer79}, \cite{ACN1}, \cite{ACN2}. The main difference here
is that supergravity is considered as part of a constrained system. This
system consists of the supergravity multiplet $\left(  e_{\mu}^{a},\psi_{\mu
},A_{\mu},u\right)  $ of $16$ bosonic and $16$ fermionic component fields and
a chiral multiplet $\left(  z,\chi,h\right)  $ of $4$ bosonic and $4$
fermionic component fields as well as the Lagrange multiplier chiral multiplet
$\left(  A+iB,\psi,l\right)  $ or linear multiplet $\left(  A,\psi,b_{\mu
}\right)  .$ This system will effectively have the same number of independent
degrees of freedom as that of a spontaneously broken supergravity multiplet
with four fermionic degrees for the massive gravitino, and four bosonic
degrees, two for the graviton and two for the scalar fields.  However, only
the model with Linear multiplet yields a constraint on the kinetic term of the
form $g^{\mu\nu}\phi_{,z\overline{z}}\partial_{\mu}z\partial_{\nu}\overline
{z}$ while the model with multiplet $S$ gives a constraint on the combination
$g^{\mu\nu}\left(  \phi_{,zz}\partial_{\mu}z\partial_{\nu}z+\phi
_{,\overline{z}\overline{z}}\partial_{\mu}\overline{z}\partial_{\nu}%
\overline{z}\right)  .$ Thus the model with coupling to the Linear multiplet
is more in line with the non-supersymmetric case. The supersymmetric case is,
however, much more complicated because of the presence of the auxiliary fields
$u,$ $A_{\mu},h,$ as these couple non-trivially to the system $\left(
S,\Sigma\right)  $ or $\left(  L,\Sigma\right)  $. Imposing and solving the
constraints coming from the Lagrange multiplier multiplet and eliminating the
auxiliary fields give rise to an action containing only the physical degrees
of freedom. For general functions $\phi\left(  z,\overline{z}\right)  $ and
$g\left(  z\right)  $ the resulting action is complicated. It should, however,
be possible to simplify the form by performing some redefinitions along the
lines of the results in \cite{Cremmer79}, \cite{Cremmer83}, \cite{ACN2}. What
is expected to be gained here is that the combination of a chiral multiplet
and a Lagrange multiplier multiplet could act as the hidden sector responsible
for supersymmetry breaking and for the appearance of mimetic dark matter in
the form of two scalar fields. These steps will be undertaken in future work
where the aim will be to present a realistic model where the Goldistino field
will be the supersymmetric partner of mimetic dark matter with details on the
local supersymmetry breaking and the actual field combinations representing
the dynamical degrees of freedom as well as the fields determining dark matter.

\section{Appendix}

In this appendix we give the full Lagrangian for supergravity with mimetic
dark matter chiral multiplet $\Sigma=\left(  z,\chi,h\right)  $ that can also
act as the hidden sector for spontaneous supersymmetry breaking. In the first
case the Lagrange multiplier multiplet is a left handed chiral multiplet
$S=\left(  A+iB,\psi,l\right)  $ embedded in a real vector multiplet. The full
Lagrangian (\ref{fullS}) is given in component form by
\begin{align}
e^{-1}L  &  =\left(  \frac{1}{3}B\phi-1\right)  \left(  \frac{1}{2}R+\frac
{1}{3}\left\vert u\right\vert ^{2}-\frac{1}{3}A_{\mu}A^{\mu}+\frac{1}{2}%
e^{-1}\overline{\psi}_{\mu}R^{\mu}\right) \nonumber\\
&  +B\phi_{,z\overline{z}}\left(  \left\vert h\right\vert ^{2}-\widehat{D}%
_{\mu}\overline{z}\widehat{D}^{\mu}z-\overline{\chi}%
\overleftrightarrow{\widehat{\slashed{D}}}\chi\right)  -B\phi_{,zz\overline
{z}}\left(  h^{\ast}\overline{\chi^{c}}\chi+\overline{\chi}\gamma^{\mu}%
\chi\widehat{D}_{\mu}z\right) \nonumber\\
&  -B\phi_{,z\overline{z}\overline{z}}\left(  h\overline{\chi}\chi
^{c}+\overline{\chi^{c}}\gamma^{\mu}\chi^{c}\widehat{D}_{\mu}\overline
{z}\right)  +B\phi_{,zz\overline{z}\overline{z}}\overline{\chi^{c}}\chi
\cdot\overline{\chi}\chi^{c}\nonumber\\
&  -\frac{1}{2}\widehat{D}^{\mu}B\widehat{D}_{\mu}\phi-\frac{i}{2}%
\widehat{D}^{\mu}A\left(  \phi_{,z}\widehat{D}_{\mu}z-\phi_{,\overline{z}%
}\widehat{D}_{\mu}\overline{z}-2\phi_{,z\overline{z}}\overline{\chi}%
\gamma_{\mu}\chi\right) \nonumber\\
&  -\frac{1}{2}\left(  \overline{l}\left(  h\phi_{,z}-\phi_{,zz}\overline
{\chi^{c}}\chi\right)  +l\left(  \overline{h}\phi_{,\overline{z}}%
-\phi_{,\overline{z}\overline{z}}\overline{\chi}\chi^{c}\right)  \right)
+i\overline{\psi}\phi_{,z\overline{z}}\left(  h^{\ast}\chi-h\chi
^{c}+\widehat{\slashed{D}}\overline{z}\chi-\widehat{\slashed{D}}z\chi
^{c}\right) \nonumber\\
&  +i\left(  \phi_{,zz\overline{z}}\overline{\chi^{c}}\chi\overline{\psi}%
\chi^{c}-\phi_{,z\overline{z}\overline{z}}\overline{\chi}\chi^{c}%
\overline{\psi}\chi\right)  -i\left(  \phi_{,z\overline{z}\overline{z}%
}\overline{\chi}\chi^{c}\overline{\chi}^{c}\psi-\phi_{,zz\overline{z}%
}\overline{\chi^{c}}\chi\overline{\chi}\psi\right) \nonumber\\
&  +\frac{i}{2}\left(  \left(  \phi_{,\overline{z}}\overline{\chi}-\phi
_{,z}\overline{\chi}^{c}\right)  \widehat{\slashed{D}}\psi-\overline{\psi
}\widehat{\slashed{D}}\left(  \phi_{,z}\chi-\phi_{,\overline{z}}\chi
^{c}\right)  \right) \nonumber\\
&  -\frac{i}{2}\overline{\psi}_{\mu}\gamma_{5}\gamma^{\mu}\left[
-iB\phi_{,z\overline{z}}\left(  \left(  h^{\ast}\chi-h\chi^{c}\right)
+\widehat{\slashed{D}}\overline{z}\chi-\widehat{\slashed{D}}z\chi^{c}\right)
\right. \nonumber\\
&  -iB\left(  \phi_{,zz\overline{z}}\left(  \overline{\chi^{c}}\chi\right)
\chi^{c}-\phi_{,z\overline{z}\overline{z}}\left(  \overline{\chi}\chi
^{c}\right)  \chi\right)  -\frac{i}{2}\widehat{\slashed{D}}B\left(  \phi
_{,z}\chi-\phi_{,\overline{z}}\chi^{c}\right) \nonumber\\
&  -\frac{1}{4}\widehat{\slashed{D}}\phi\psi-\frac{1}{2}\gamma_{5}\gamma^{\mu
}\left(  \phi_{,z}\chi-\phi_{,\overline{z}}\chi^{c}\right)  \widehat{D}_{\mu
}A-\frac{1}{2}\widehat{\left(  \phi_{,\overline{z}}\overline{h}-\phi
_{,zz}\overline{\chi^{c}}\chi\right)  }\psi\nonumber\\
&  \left.  -\frac{1}{4}\gamma_{5}\gamma^{\mu}\psi\left(  \phi_{,z}%
\widehat{D}_{\mu}z-\phi_{,\overline{z}}\widehat{D}_{\mu}\overline{z}%
-2\phi_{,z\overline{z}}\overline{\chi}\gamma_{\mu}\chi\right)  +\frac{i}%
{2}\widehat{\overline{l}}\left(  \phi_{,z}\chi-\phi_{,\overline{z}}\chi
^{c}\right)  \right] \nonumber\\
&  +\frac{2}{3}\left(  A^{\mu}+\frac{3i}{8}e^{-1}\epsilon^{\mu\rho\sigma\tau
}\overline{\psi}_{\rho}\gamma_{\tau}\psi_{\sigma}\right)  \left[  \frac{i}%
{2}B\left(  \phi_{,z}\widehat{D}_{\mu}z-\phi_{,\overline{z}}\widehat{D}_{\mu
}\overline{z}-2\phi_{,z\overline{z}}\overline{\chi}\gamma_{\mu}\chi\right)
\right. \nonumber\\
&  \left.  +\frac{1}{2}\phi\widehat{D}_{\mu}A+\frac{1}{2}\overline{\psi}%
\gamma_{\mu}\gamma_{5}\left(  \phi_{,z}\chi-\phi_{,\overline{z}}\chi
^{c}\right)  \right] \nonumber\\
&  +\frac{i}{3}e^{-1}\left(  -iB\left(  \phi_{,\overline{z}}\overline{\chi
}-\phi_{,z}\overline{\chi}^{c}\right)  +\frac{1}{2}\phi\overline{\psi}\right)
\gamma_{5}\gamma_{\mu}R^{\mu}\nonumber\\
&  +\frac{i}{8}e^{-1}\epsilon^{\mu\nu\rho\sigma}\overline{\psi}_{\mu}%
\gamma_{\nu}\psi_{\rho}\left(  -iB\left(  \phi_{,\overline{z}}\overline{\chi
}-\phi_{,z}\overline{\chi}^{c}\right)  +\frac{1}{2}\phi\overline{\psi}\right)
\psi_{\sigma}\nonumber\\
&  +\frac{1}{3}\left(  \overline{u}\left(  -B\left(  h\phi_{,z}-\phi
_{,zz}\overline{\chi^{c}}\chi\right)  +\frac{1}{2}\phi l-\frac{i}{2}%
\overline{\psi}_{R}\left(  \phi_{,z}\chi_{L}-\phi_{,\overline{z}}\chi_{L}%
^{c}\right)  +\frac{i}{2}\left(  \phi_{,\overline{z}}\overline{\chi}_{R}%
-\phi_{,z}\overline{\chi^{c}}_{R}\right)  \psi_{L}\right)  +h.c\right)
\nonumber\\
&  +\frac{1}{2}\left(  g_{,z}h-g_{,,zz}\overline{\chi}\chi+ug+g_{,z}%
\overline{\psi}_{\mu}\gamma^{\mu}\chi+g\overline{\psi}_{\mu}\sigma^{\mu\nu
}\psi_{\nu R}+h.c\right)  \label{LagrangianA}%
\end{align}

The second Lagrangian corresponds to the case where the Lagrange multiplier
supermultiplet is a linear multiplet $L=\left(  A,\psi,b\right)  $ embedded in
a real vector multiplet. When expressed in terms of components the full
Lagrangian $\left(  \ref{FullB}\right)  $ becomes
\begin{align}
e^{-1}L  &  =\left(  \frac{1}{3}A\phi-1\right)  \left(  \frac{1}{2}R+\frac
{1}{3}\left\vert u\right\vert ^{2}-\frac{1}{3}A_{\mu}A^{\mu}+\frac{1}{2}%
e^{-1}\overline{\psi}_{\mu}R^{\mu}\right) \nonumber\\
&  +A\phi_{,z\overline{z}}\left(  \left\vert h\right\vert ^{2}-\widehat{D}%
_{\mu}\overline{z}\widehat{D}^{\mu}z-\overline{\chi}%
\overleftrightarrow{\widehat{\slashed{D}}}\chi\right)  -A\phi_{,zz\overline
{z}}\left(  \overline{h}\overline{\chi^{c}}\chi+\overline{\chi}\gamma^{\mu
}\chi\widehat{D}_{\mu}z\right) \nonumber\\
&  -A\phi_{,z\overline{z}\overline{z}}\left(  h\overline{\chi}\chi
^{c}+\overline{\chi^{c}}\gamma^{\mu}\chi^{c}\widehat{D}_{\mu}\overline
{z}\right)  +A\phi_{,zz\overline{z}\overline{z}}\left(  \overline{\chi^{c}%
}\chi\right)  \left(  \overline{\chi}\chi^{c}\right)  -\frac{1}{2}\phi
\square^{P}A-\frac{1}{2}\widehat{D}^{\mu}A\widehat{D}_{\mu}\phi\nonumber\\
&  -\frac{i}{2}b^{\mu}\left(  \phi_{,z}\widehat{D}_{\mu}z-\phi_{,\overline{z}%
}\widehat{D}_{\mu}\overline{z}-2\phi_{,z\overline{z}}\overline{\chi}\gamma
_{5}\gamma_{\mu}\chi\right)  +i\phi_{,z\overline{z}}\overline{\psi}\left(
\overline{h}\chi-h\chi^{c}+\widehat{\slashed{D}}\overline{z}\chi
-\widehat{\slashed{D}}z\chi^{c}\right) \nonumber\\
&  -i\phi_{,z\overline{z}}\left(  h\overline{\chi}-\overline{h}\overline
{\chi^{c}}+\widehat{\slashed{D}}z\overline{\chi}-\widehat{\slashed{D}}%
\overline{z}\overline{\chi^{c}}\right)  \psi+\frac{i}{2}\left(  \phi
_{,\overline{z}}\overline{\chi}-\phi_{,z}\overline{\chi^{c}}\right)
\widehat{\slashed{D}}\psi\nonumber\\
&  -\frac{i}{2}\overline{\psi}_{\mu}\gamma_{5}\gamma^{\mu}\left[
A\phi_{,z\overline{z}}\left(  -i\overline{h}\chi+ih\chi^{c}%
-i\widehat{\slashed{D}}\overline{z}\chi+i\widehat{\slashed{D}}z\chi
^{c}\right)  -\frac{1}{2}\phi\widehat{\slashed{D}}\psi\right. \nonumber\\
&  -\frac{i}{2}\widehat{\slashed{D}}A\left(  \phi_{,z}\chi-\phi_{,\overline
{z}}\chi^{c}\right)  -\frac{1}{4}\widehat{\slashed{D}}\phi\psi-iA\left(
\phi_{,zz\overline{z}}\overline{\chi^{c}}\chi\chi^{c}-\phi_{,z\overline
{z}\overline{z}}\overline{\chi}\chi^{c}\chi\right) \nonumber\\
&  \left.  -\frac{1}{2}\gamma_{5}\gamma^{\mu}\left(  b_{\mu}\left(  \phi
_{,z}\chi-\phi_{,\overline{z}}\chi^{c}\right)  +\frac{1}{2}\left(  \phi
_{,z}\widehat{D}_{\mu}z-\phi_{,\overline{z}}\widehat{D}_{\mu}\overline
{z}-2\phi_{,z\overline{z}}\overline{\chi}\gamma_{\mu}\chi\right)  \psi\right)
-\widehat{\frac{1}{2}\left(  \phi_{,\overline{z}}\overline{h}-\phi
_{,\overline{z}\overline{z}}\overline{\chi}\chi^{c}\right)  }\psi\right]
\nonumber\\
&  +\frac{2}{3}\left(  A^{\mu}+\frac{3i}{8}e^{-1}\epsilon^{\mu\rho\sigma\tau
}\overline{\psi}_{\rho}\gamma_{\tau}\psi_{\sigma}\right)  \left[  \frac{i}%
{2}A\left(  \phi_{,z}\widehat{D}_{\mu}z-\phi_{,\overline{z}}\widehat{D}_{\mu
}\overline{z}-2\phi_{,z\overline{z}}\overline{\chi}\gamma_{\mu}\chi\right)
\right. \nonumber\\
&  \qquad\hspace{2in}\qquad\left.  +\frac{1}{2}\phi b_{\mu}+\frac{1}%
{2}\overline{\psi}\gamma_{\mu}\gamma_{5}\left(  \phi_{,z}\chi-\phi
_{,\overline{z}}\chi^{c}\right)  \right] \nonumber\\
&  +\frac{i}{3}e^{-1}\left(  -iA\left(  \phi_{,\overline{z}}\overline{\chi
}-\phi_{,\overline{z}}\overline{\chi}^{c}\right)  +\frac{1}{2}\phi
\overline{\psi}\right)  \gamma_{5}\gamma_{\mu}R^{\mu}\nonumber\\
&  +\frac{i}{8}e^{-1}\epsilon^{\mu\nu\rho\sigma}\overline{\psi}_{\mu}%
\gamma_{\nu}\psi_{\rho}\left(  -iA\left(  \phi_{,\overline{z}}\overline{\chi
}-\phi_{,\overline{z}}\overline{\chi}^{c}\right)  +\frac{1}{2}\phi
\overline{\psi}\right)  \psi_{\sigma}\nonumber\\
&  -\frac{1}{3}\left(  \overline{u}\left(  -A\left(  \phi_{,z}h-\phi
_{,zz}\overline{\chi^{c}}\chi\right)  -\frac{i}{2}\overline{\psi}_{R}\left(
\phi_{,z}\chi_{L}-\phi_{,\overline{z}}\chi_{L}^{c}\right)  +\frac{i}{2}\left(
\phi_{,\overline{z}}\overline{\chi}_{R}-\phi_{,z}\overline{\chi^{c}}%
_{R}\right)  \psi_{L}\right)  +h.c\right) \nonumber\\
&  +\frac{1}{2}\left(  g_{,z}h-g_{,,zz}\overline{\chi}\chi+ug+g_{,z}%
\overline{\psi}_{\mu}\gamma^{\mu}\chi+g\overline{\psi}_{\mu}\sigma^{\mu\nu
}\psi_{\nu R}+h.c\right)  \label{LagrangianB}%
\end{align}

\begin{acknowledgement}
This research is supported in parts by the National Science Foundation Grant
No. Phys-1912998.
\end{acknowledgement}


\begin{thebibliography}{99}                                                                                               %


\bibitem {mimetic}A. H. Chamseddine and V. Mukhanov, \textit{Mimetic Dark
Matter, }JHEP \textbf{1311 }(2013) 135.

\bibitem {Golovnev}A. Golovnev, \textit{On the Recently Proposed Mimetic Dark
Matter, }Phys. Lett. \textbf{B728 }(2014) 39.

\bibitem {mimcos}A. H. Chamseddine, V. Mukhanov and A. Vikman,
\textit{Cosmology with Mimetic Matter, }JCAP \textbf{1406 }(2014) 017.

\bibitem {Singular}A. H. Chamseddine and V. Mukhanov, \textit{Resolving
Cosmological Singularities, }JCAP \textbf{1703}

\bibitem {BH}A. H. Chamseddine and V. Mukhanov, \textit{Nonsingular Black
Hole, }Eur. Phys. J. \textbf{C77} (2017) 83.

\bibitem {AF}A. H. Chamseddine, V. Mukhanov and T. Russ,
\textit{Asymptotically Free Mimetic Gravity, }Eur. Phys. J. \textbf{C79
}(2019) 558.

\bibitem {minimal}A. H. Chamseddine, V. Mukhanov and T. Russ, \textit{Black
hole remnants, }JHEP (2019) 104.

\bibitem {Cremmer79}E. Cremmer, B. Julia, J. Scherk, S. Ferrara, L. Girardello
and P. van Niewenhuizen, \textit{Spontaneous symmetry breaking and Higgs
effect in supergravity without cosmological constant, }Nucl. Phys.
B\textbf{147 }(1979) 105.

\bibitem {Stelle}K. Stelle and P. C. West, \textit{Relation between vector and
scalar multiplets and gauge invariance in supergravity, }Nucl. Phys.
B\textbf{145 }(1978) 175.

\bibitem {Cremmer 82}E. Cremmer, S. Ferrara, L. Girardello and A. van Proeyen,
\textit{Coupling supersymmetric Yang-Mills theories to supergravity, }Phys.
Lett. \textbf{116}B, (1982) 231.

\bibitem {ACN1}A. H. Chamseddine, R. Arnowitt and P. Nath, \textit{Locally
supersymmetric grand unification, }Phys. Rev. Lett. \textbf{49 }(1982) 970.

\bibitem {Cremmer83}E. Cremmer, S. Ferrara, L. Girardello and A. van Proeyen,
\textit{Yang-Mills theories with local supersymmetry: Lagrangian,
transformation laws and super-Higgs effect, }Nucl. Phys. B\textbf{212} (1983)
413.\textbf{ }(2017) 009.

\bibitem {ACN2}P. Nath, R. Arnowitt and A. H. Chamseddine, \textit{Applied N=1
Supergravity, }World Scientific, 1984.

\bibitem {van}P. van Niewenhuizen, \textit{Lectures in supergravity theory,
}in Recent developments in gravitation, Cargese 1978, editors M. Levy and S.
Deser, Plenum Press 1979.

\bibitem {inhomogeneous}A. H. Chamseddine and V. Mukhanov,
\textit{Inhomogeneous dark energy, }JCAP \textbf{02 }(2016) 040.

\bibitem {Ola}O. Malaeb, \textit{Hamiltonian formulation of mimetic gravity,
}Phys. Rev. D\textbf{19 }(2015) 10.

\bibitem {derendinger}C. Aulakh, J. -P. Derendinger and S. Ouvry, \textit{On
the reducibility of }$16+16$ \textit{supergravity, }Phys. Lett. B\textbf{169
}(1986) 201.

\bibitem {Horava}A. H. Chamseddine, V. Mukhanov and T. Russ, \textit{Mimetic
Horava gravity, }Phys. Lett. B\textbf{798 }(2019) 134939.
\end{thebibliography}
\end{document}